\documentclass[11pt]{article}
\usepackage{amssymb,amsmath,amsfonts}
\usepackage{graphicx}
\usepackage{graphics}
\usepackage{eepic,epsfig}

\makeatletter
\@addtoreset{equation}{section}

\makeatother

\begin{document}

\title{Fermionic currents in AdS spacetime with compact dimensions}
\author{ S. Bellucci$^{1}$\thanks{%
E-mail: bellucci@lnf.infn.it }, A. A. Saharian$^{2}$\thanks{%
E-mail: saharian@ysu.am }, V. Vardanyan$^{2,3,4}$\thanks{%
E-mail: vardanyan@lorentz.leidenuniv.nl } \vspace{0.3cm} \\
%EndAName
\textit{$^1$ INFN, Laboratori Nazionali di Frascati,}\\
\textit{Via Enrico Fermi 40,00044 Frascati, Italy} \vspace{0.3cm}\\
\textit{$^2$ Department of Physics, Yerevan State University,}\\
\textit{1 Alex Manoogian Street, 0025 Yerevan, Armenia }\vspace{0.3cm}\\
\textit{$^3$ Lorentz Institute for Theoretical Physics, Leiden University, }%
\\
\textit{2333 CA Leiden, The Netherlands }\vspace{0.3cm}\\
\textit{$^4$ Leiden Observatory, Leiden University, 2300 RA Leiden, The
Netherlands}}
\maketitle

\begin{abstract}
We derive a closed expression for the vacuum expectation value (VEV) of the
fermionic current density in a $(D+1)$-dimensional locally AdS spacetime
with an arbitrary number of toroidally compactified Poincar\'{e} spatial
dimensions and in the presence of a constant gauge field. The latter can be
formally interpreted in terms of a magnetic flux treading the compact
dimensions. In the compact subspace, the field operator obeys
quasiperiodicity conditions with arbitrary phases. The VEV of the charge
density is zero and the current density has nonzero components along the
compact dimensions only. They are periodic functions of the magnetic flux
with the period equal to the flux quantum and tend to zero on the AdS
boundary. Near the horizon, the effect of the background gravitational field
is small and the leading term in the corresponding asymptotic expansion
coincides with the VEV for a massless field in the locally Minkowski bulk.
Unlike the Minkowskian case, in the system consisting an equal number of
fermionic and scalar degrees of freedom, with same masses, charges and
phases in the periodicity conditions, the total current density does not
vanish. In these systems, the leading divergences in the scalar and
fermionic contributions on the horizon are canceled and, as a consequence of
that, the charge flux, integrated over the coordinate perpendicular to the
AdS boundary, becomes finite. We show that in odd spacetime dimensions the
fermionic fields realizing two inequivalent representations of the Clifford
algebra and having equal phases in the periodicity conditions give the same
contribution to the VEV of the current density. Combining the contributions
from these fields, the current density in odd-dimensional $C$-,$P$- and $T$%
-symmetric models are obtained. As an application, we consider the ground
state current density in curved carbon nanotubes described in terms of a
(2+1)-dimensional effective Dirac model.
\end{abstract}


\bigskip

\section{Introduction}

\label{sec:introd}

In a number of physical problems one needs to consider the model in
background of manifolds with compact subspaces. The presence of extra
compact dimensions is an inherent feature of fundamental theories unifying
physical interactions, like Kaluza-Klein theories, supergravity and string
theories. The compact spatial dimensions also appear in the low-energy
effective description of some condensed matter systems. Examples for the
latter are the cylindrical and toroidal carbon nanotubes and topological
insulators.

In quantum field theory, the periodicity conditions imposed along the
compact dimensions modify the spectrum of the zero-point fluctuations of
quantum fields. As a consequence of that, the vacuum expectation values
(VEVs) of physical quantities are shifted by an amount depending on the
geometry of the compact subspace. This general phenomenon, induced by the
nontrivial topology, is the analog of the Casimir effect (for reviews see
Ref. \cite{Most97}) where the change in the spectrum of the vacuum
fluctuations is caused by the presence of boundaries (conductors,
dielectrics, branes in braneworld scenarios, etc.). It is known as the
topological Casimir effect and has been investigated for different fields,
bulk geometries and topologies. The corresponding vacuum energy depends on
the lengths of the compact dimensions and the topological Casimir effect has
been considered as a stabilization mechanism for the moduli fields related
to extra dimensions. In addition, the vacuum energy induced by the
compactification of spatial dimensions can serve as a model for the dark
energy driving the accelerated expansion of the universe at a recent epoch
\cite{Milt03}.

For charged quantum fields, important characteristics for a given state are
the expectation values of the charge and current densities. In the present
paper we investigate the VEV\ of the current density for a massive fermionic
field in the background of a locally anti-de Sitter (AdS) spacetime with an
arbitrary number of toroidally compactified spatial dimensions (for a
discussion of physical effects in models with toroidal dimensions, see for
instance, Ref. \cite{Eliz95}). The corresponding problem for a scalar field
with general coupling to the Ricci scalar has been previously considered in
Ref. \cite{Beze15} (see also Refs. \cite{Bell15b,Bell16} for additional
effects induced by the presence of branes). Both the zero and finite
temperature expectation values of the current density for charged scalar and
fermionic fields in the background of flat spacetime with toral dimensions
were investigated in Refs. \cite{Beze13c,Bell10}. The results were applied
to the electronic subsystem of cylindrical and toroidal carbon nanotubes
described in terms of a $(2+1)$-dimensional effective field theory. The
vacuum current densities for charged scalar and Dirac spinor fields in de
Sitter spacetime with toroidally compact spatial dimensions are considered
in Ref. \cite{Bell13b}. The influence of boundaries on the vacuum currents
in topologically nontrivial spaces are studied in Refs. \cite%
{Bell13,Bell15,Bell16b} for scalar and fermionic fields. The effects of
nontrivial topology induced by the compactification of a cosmic string along
its axis have been discussed in Ref. \cite{Beze13}. The vacuum energy and
the VEV of the energy-momentum tensor in AdS spacetime with compact
subspaces were investigated in Ref. \cite{Flac03}.

Our choice of AdS spacetime as the background geometry is motivated by its
importance in several recent developments of quantum field theory, gravity
and condensed matter physics. The early interest to AdS spacetime as a bulk
geometry in quantum field theory was motivated by principal questions of the
quantization procedure on curved backgrounds. Because of the maximal
symmetry of AdS spacetime, this procedure can be realized explicitly.
Compared to the case of the Minkowski bulk, here principally new features
arise related to the lack of global hyperbolicity and the presence of both
regular and irregular modes. The main reason of the lack of hyperbolicity is
that the AdS spacetime possesses a timelike boundary at spatial infinity
through which the information may be lost or gained in finite coordinate
time \cite{Avis78}. As a consequence, boundary conditions should be imposed
at infinity to ensure a consistent quantum field theory. The natural
appearance of AdS spacetime as a ground state in certain supergravity
theories and also as the near horizon geometry of the extremal black holes
and domain walls has stimulated further interest in quantum fields
propagating on that background. Moreover, the AdS spacetime is a constant
negative curvature manifold and the corresponding length scale can be used
for the regularization of infrared divergences in interacting quantum field
theories \cite{Call90}. The dimension of the AdS isometry group is the same
as that of the Poincar\'{e} group and the regularization is realized without
reducing the symmetries.

The renewed interest in physical models on AdS bulk is closely related to
two rapidly developing fields in theoretical physics: the gauge/gravity
duality and the braneworld scenario. The AdS spacetime played a crucial role
in the original formulations of both these concepts in the form of the
AdS/CFT correspondence \cite{Mald98} and the Randall-Sundrum type
braneworlds \cite{Rand99}. The AdS/CFT correspondence (for reviews see Ref.
\cite{Ahar00}) is a type of holographic duality between two theories living
in spacetimes with different dimensions: string theories or supergravity in
the AdS bulk from one side and a conformal field theory localized on the AdS
boundary from another. Among many interesting consequences, this duality
opens an important opportunity to study quantum field theoretical effects in
strongly coupled regime using a classical gravitational theory. It has also
been used for the investigation of non-equilibrium phenomena in strongly
coupled condensed matter systems. One of the applications is the holographic
model for superconductors suggested in Ref. \cite{Hart08} (for a recent
discussion with references, see e.g., Ref. \cite{Cai15}). In this model, the
quantum physics of strongly correlated condensed matter system is mapped to
the gravitational dynamics with black holes in one higher dimension.

The braneworld scenario (see Ref. \cite{Maar10} for a review) provides an
interesting alternative to the standard Kaluza-Klein compactification of
extra dimensions. It uses the concept of brane as a submanifold embedded in
a higher dimensional spacetime, on which the standard model fields are
confined. Braneworlds naturally appear in the string/M-theory context and
provide interesting possibilities to solve or to address from a different
point of view various problems in cosmology and particle physics. In the
model introduced by Randall and Sundrum the background geometry consists of
two parallel branes, with positive and negative tensions, embedded in a five
dimensional AdS bulk \cite{Rand99}. The fifth coordinate is compactified on
orbofold and the branes are located at the two fixed points. The large
hierarchy between the Planck and electroweak mass scales is generated by the
large physical volume of the extra dimension. From the point of view of
embedding the corresponding models into a more fundamental theory, such as
string/M-theory, one may expect the presence of additional extra dimensions
compactified on an internal manifold. Here we will consider a simple case of
toroidal compactification of spatial dimensions in Poincar\'{e} coordinates.

The plan of the paper is as follows. In the next section the problem is
formulated and the complete set of fermionic modes is presented. By using
these modes, in Section \ref{sec:Current}, we evaluate the VEV of the
fermionic current density along compact dimensions. The asymptotics near the
AdS boundary and near the horizon are investigated and limiting cases are
discussed. In Section \ref{sec:OddDim}, we consider the current density in $%
C $-, $P$-, $T$-symmetric odd-dimensional models. The corresponding VEV is
obtained by combining the results for two fermionic fields realizing the
irreducible representations of the Clifford algebra. Applications are given
to graphene made structures realizing the geometry under consideration. The
main results of the paper are summarized in Section \ref{sec:Conc}. In
Appendix \ref{sec:Repr2} we consider an alternative representation of the
Dirac matrices allowing for the separation of the equations for the upper
and lower components of the fermionic mode functions in Poincar\'{e}
coordinates. The main steps for the evaluation of the mode-sum for the
current density are presented in Appendix \ref{sec:App}.

\section{Problem setup and the fermionic modes}

\label{sec:Setup}

The dynamics of a fermionic field $\psi (x)$ in a $(D+1)$-dimensional curved
background with a metric tensor $g_{\mu \nu }(x)$ and in the presence of an
abelian gauge field $A_{\mu }(x)$ is described by the Dirac equation
\begin{equation}
i\gamma ^{\mu }D_{\mu }\psi -m\psi =0,  \label{eom}
\end{equation}%
with the gauge extended covariant derivative $D_{\mu }=\partial _{\mu
}+\Gamma _{\mu }+ieA_{\mu }$. Here, $\Gamma _{\mu }$ is the spin connection
and the curved spacetime Dirac matrices $\gamma ^{\mu }$ are expressed in
terms of the corresponding flat spacetime matrices $\gamma ^{(b)}$ as $%
\gamma ^{\mu }=e_{(b)}^{\mu }\gamma ^{(b)}$, where $e_{(b)}^{\mu }$ are the
tetrad fields. In this and in the next sections we consider a fermionic
field realizing the irreducible representation of the Clifford algebra. In $%
(D+1)$-dimensional spacetime the corresponding Dirac matrices are $N\times N$
matrices, where $N=2^{[(D+1)/2]}$ and $[x]$ is the integer part of $x$ (for
the Dirac matrices in an arbitrary number of the spacetime dimension see,
for example, Ref. \cite{Park09}). In even-dimensional spacetimes the
irreducible representation is unique up to a similarity transformation,
whereas in odd-dimensional spacetimes there are two inequivalent irreducible
representations (see Section \ref{sec:OddDim} below).

As a background geometry we consider a locally AdS spacetime with the line
element
\begin{equation}
ds^{2}=e^{-2y/a}\eta _{ik}dx^{i}dx^{k}-dy^{2},  \label{ads}
\end{equation}%
where $a$ is the curvature radius, $i,k=0,1,\ldots ,D-1$ and $\eta _{ik}=%
\mathrm{diag}(1,-1,\ldots ,-1)$. We assume that the subspace covered by the
coordinates $(x^{p+1},\ldots ,x^{D-1})$ is compactified to a $q$-dimensional
torus $T^{q}$ with $q=D-p-1$. The length of the $l$th compact dimension will
be denoted by $L_{l}$, $0\leqslant x^{l}\leqslant L_{l}$, $l=p+1,\ldots ,D-1$%
. The subspace $(x^{1},\ldots ,x^{p})$ has trivial topology $R^{p}$ with $%
-\infty <x^{l}<+\infty $, $l=1,\ldots ,p$, and for the coordinate $y$ one
has $-\infty <y<+\infty $. Note that the compactification to the torus does
not change the local AdS geometry with the scalar curvature $R=-D(D+1)/a^{2}$%
. In terms of a new spatial coordinate $z=ae^{y/a}$, $0\leqslant z<\infty $,
the line element is presented in a conformally-flat form
\begin{equation}
ds^{2}=(a/z)^{2}(\eta _{ik}dx^{i}dx^{k}-dz^{2}).  \label{ads_z_coord}
\end{equation}%
The AdS boundary and horizon correspond to the hypersurfaces $z=0$ and $%
z=\infty $, respectively. Note that $L_{l}$ is the coordinate length of the
compact dimension. For a given $z$, the proper length is given by $%
L_{(p)l}=(a/z)L_{l}$. The latter decreases with increasing $z$. In the
conformal coordinates $(x^{1},\ldots ,x^{D-1},x^{D}=z)$, the tetrad fields
can be chosen as $e_{(b)}^{\mu }=(z/a)\delta _{b}^{\mu }$. For the
corresponding components of the spin connection one gets $\Gamma _{k}=\eta
_{kl}\gamma ^{(D)}\gamma ^{(l)}/(2z)$ for $k=0,\ldots ,D-1$, and $\Gamma
_{D}=0$.

In the presence of compact dimensions, the field equation (\ref{eom}) should
be supplemented by the periodicity conditions on the field operator along
those directions. Here, we will impose quasiperiodicity conditions
\begin{equation}
\psi (t,x^{1},\ldots ,x^{l}+L_{l},\ldots ,x^{D})=e^{i\alpha _{l}}\psi
(t,x^{1},\ldots ,x^{l},\ldots ,x^{D}),  \label{PerCond1}
\end{equation}%
with constant phases $\alpha _{l}$, $l=p+1,\ldots ,D-1$. The special cases $%
\alpha _{l}=0$ and $\alpha _{l}=\pi $ correspond to untwisted and twisted
fermionic fields, most frequently discussed in the literature. For the gauge
field we will consider the simplest configuration $A_{\mu }=\mathrm{const}$.
Though the corresponding field tensor vanishes, the nontrivial topology of
the background spacetime gives rise the Aharonov-Bohm like effect on the
VEVs of physical observables. For this special field configuration, the
gauge field can be removed from the field equation by the gauge
transformation
\begin{equation}
\psi =\psi ^{\prime }e^{-ie\chi },\;A_{\mu }=A_{\mu }^{\prime }+\partial
_{\mu }\chi ,  \label{gauge}
\end{equation}%
with the function $\chi =A_{\mu }x^{\mu }$. In the new gauge $A_{\mu
}^{\prime }=0$. However, the gauge potential does not disappear from the
problem completely. The gauge transformation of the field operator modifies
the corresponding periodicity conditions. For the new field it takes the
form
\begin{equation}
\psi ^{\prime }(t,x^{1},\ldots ,x^{l}+L_{l},\ldots ,x^{D})=e^{i\tilde{\alpha
_{l}}}\psi ^{\prime }(t,x^{1},\ldots ,x^{l},\ldots ,x^{D}),  \label{PerCond2}
\end{equation}%
with the new phases
\begin{equation}
\tilde{\alpha _{l}}=\alpha _{l}+eA_{l}L_{l}.  \label{alftild}
\end{equation}%
The VEVs of physical observables will depend on the set $\{\alpha
_{l},A_{l}\}$ in the form of the combination (\ref{alftild}). Under the
gauge transformation (\ref{gauge}) with $\chi =b_{\mu }x^{\mu }$ and
constant $b_{\mu }$, we obtain a new set $\{\alpha
_{l}+eb_{l}L_{l},A_{l}-b_{l}\}$. However, the combination (\ref{alftild})
remains invariant. Note that the phase shift in Eq. (\ref{alftild}), induced
by the gauge field, can be presented as $eA_{l}L_{l}=-2\pi \Phi _{l}/\Phi
_{0}$, where $\Phi _{0}=2\pi /e$ is the flux quantum. The quantity $\Phi
_{l} $ can be formally interpreted as the magnetic flux enclosed by the $l$%
th compact dimension, $\Phi _{l}=-\oint dx^{l}A_{l}$ (no summation over $l$%
). This flux acquires real physical meaning in models where the spacetime
under consideration is embedded in a higher-dimensional manifold as a
hypersurface (like branes in braneworld scenario) on which the fermionic
field $\psi (x)$ is localized. In what follows we will work in the gauge $%
(A_{\mu }^{\prime }=0,\psi ^{\prime }(x))$ omitting the prime for the
fermionic field. In this gauge, in Eq. (\ref{eom}) we have $D_{\mu
}=\partial _{\mu }+\Gamma _{\mu }$. Of course, the VEVs of physical
observables do not depend on the choice of the gauge.

We are interested in the VEV of the fermionic current density $\langle
0|j^{\mu }(x)|0\rangle \equiv \langle j^{\mu }(x)\rangle $ with the operator
$j^{\mu }(x)=e\bar{\psi}(x)\gamma ^{\mu }\psi (x)$, where $|0\rangle $
stands for the vacuum state and $\bar{\psi}(x)=\psi ^{\dagger }\gamma ^{(0)}$%
. The VEV is presented as the coincidence limit
\begin{equation}
\langle j^{\mu }(x)\rangle =-\frac{e}{2}\lim_{x^{\prime }\rightarrow x}%
\mathrm{Tr}(\gamma ^{\mu }S^{(1)}(x,x^{\prime })),  \label{VEVj}
\end{equation}%
with the two-point function $S_{ik}^{(1)}(x,x^{\prime })=\langle 0|[\psi
_{i}(x),\bar{\psi}_{k}(x^{\prime })]|0\rangle $ and with $i$, $k$ being
spinor indices. Expanding the field operator in terms of a complete set of
mode functions $\{\psi _{\beta }^{(+)}(x),\psi _{\beta }^{(-)}(x)\}$ for the
Dirac equation and using the anticommutation relations for the annihilation
and creation operators, one gets the mode-sum representation%
\begin{equation}
\langle j^{\mu }\rangle =\frac{e}{2}\sum_{\beta }\left[ \bar{\psi}_{\beta
}^{(-)}(x)\gamma ^{\mu }\psi _{\beta }^{(-)}(x)-\bar{\psi}_{\beta
}^{(+)}(x)\gamma ^{\mu }\psi _{\beta }^{(+)}(x)\right] .  \label{current}
\end{equation}%
Here, $\beta $ is the set of quantum numbers specifying the fermionic modes
and $\sum_{\beta }$ is understood as summation for discrete subset of
quantum numbers and integration over the continues ones. Following Ref. \cite%
{Eliz13} (see also \cite{Beze13b} for the geometry with a cosmic string
perpendicular to the AdS boundary), in the discussion below we will take the
Dirac matrices in the representation%
\begin{equation}
\gamma ^{0}=i\frac{z}{a}\left(
\begin{array}{cc}
0 & -1 \\
1 & 0%
\end{array}%
\right) ,\;\gamma ^{l}=i\frac{z}{a}\left(
\begin{array}{cc}
-\sigma _{l} & 0 \\
0 & \sigma _{l}%
\end{array}%
\right) ,  \label{gamma}
\end{equation}%
with $l=1,\ldots ,D$. The $N/2\times N/2$ matrices $\sigma _{l}$ obey the
anticommutation relations $\sigma _{l}\sigma _{k}+\sigma _{k}\sigma
_{l}=2\delta _{lk}$. For hermitian $\sigma _{l}$ one has $\gamma ^{0\dagger
}=\gamma ^{0}$ and $\gamma ^{l\dagger }=-\gamma ^{l}$.

The complete set of the modes for the problem under consideration can be
found in the way similar to that used in Ref. \cite{Eliz13} for the usual
AdS bulk. The positive-energy modes are presented as%
\begin{equation}
\psi _{\beta }^{(+)}=C_{\beta }^{(+)}z^{\frac{D+1}{2}}e^{i\mathbf{kx}%
-i\omega t}\left(
\begin{array}{c}
\widehat{Z}_{-}(\lambda z)w^{(\sigma )} \\
\frac{1}{\omega }\widehat{Z}_{+}(\lambda z)\left( i\lambda +\mathbf{k\sigma }%
\right) w^{(\sigma )}%
\end{array}%
\right) ,  \label{psi+}
\end{equation}%
where $0\leqslant \lambda <\infty $, $\omega =\sqrt{\lambda ^{2}+k^{2}}$, $%
\mathbf{kx}=\sum_{l=1}^{D-1}k_{l}x^{l}$, $\mathbf{k\sigma }%
=\sum_{l=1}^{D-1}k_{l}\sigma _{l}$, and
\begin{equation}
\widehat{Z}_{\pm }(x)=\left(
\begin{array}{cc}
J_{ma\pm 1/2}(x) & 0 \\
0 & J_{ma\mp 1/2}(x)%
\end{array}%
\right) ,  \label{Zipm}
\end{equation}%
with $J_{\nu }(x)$ being the Bessel function. In Eq. (\ref{psi+}), $%
w^{(\sigma )}$, $\sigma =$ $1,\ldots ,N/2$, are one-column matrices having $%
N/2$ rows, with the elements $w_{l}^{(\sigma )}=\delta _{l\sigma }$. For the
negative-energy modes we get
\begin{equation}
\psi _{\beta }^{(-)}=C_{\beta }^{(-)}z^{\frac{D+1}{2}}e^{i\mathbf{kx}%
-i\omega t}\left(
\begin{array}{c}
\frac{1}{\omega }\widehat{Z}_{-}(\lambda z)\left( i\lambda -\mathbf{k\sigma }%
\right) w^{(\sigma )} \\
\widehat{Z}_{+}(\lambda z)w^{(\sigma )}%
\end{array}%
\right) .  \label{psi-}
\end{equation}%
The fermionic states are now specified by the set of quantum numbers $\beta
=(\mathbf{k},\lambda ,\sigma )$.

The eigenvalues of the momentum components along the compact dimensions are
found from the quasiperiodicity conditions (\ref{PerCond2}):%
\begin{equation}
k_{l}=\frac{2\pi n_{l}+\tilde{\alpha _{l}}}{L_{l}},\;l=p+1,\ldots ,D-1,
\label{kl}
\end{equation}%
with $n_{l}=0,\pm 1,\pm 2,\ldots $. For the components of the momentum along
the noncompact dimensions one has $-\infty <k_{l}<+\infty $, $l=1,\ldots ,p$%
. The coefficients $C_{\beta }^{(\pm )}$ are found from the normalization
condition $\int d^{D}x\,(a/z)^{D}\psi _{\beta }^{(\pm )\dagger }\psi _{\beta
^{\prime }}^{(\pm )}=\delta _{\beta \beta ^{\prime }}$, where $\delta
_{\beta \beta ^{\prime }}$ is understood as the Kronecker delta for discrete
quantum numbers and the Dirac delta function for the continuous ones. From
that condition one gets
\begin{equation}
\left\vert C_{\beta }^{(\pm )}\right\vert ^{2}=\frac{\lambda }{2(2\pi
)^{p}V_{q}a^{D}},  \label{Cbet}
\end{equation}%
where $V_{q}=L_{p+1}\cdots L_{D-1}$ is the volume of the compact subspace.

In defining the mode functions (\ref{psi+}) and (\ref{psi-}), as a solution
of the Bessel equation we have taken the Bessel function of the first kind
(see Eq. (\ref{Zipm})). In the range $ma\geqslant 1/2$ this choice is
uniquely dictated by the normalizability condition of the mode functions:
for the solution with the Neumann function $Y_{ma\pm 1/2}(x)$ the
normalization integral over $z$ diverges at the lower limit $z=0$. For $%
ma<1/2$ both solutions of the Bessel equation are normalizable and, in
general, we can take in Eq. (\ref{Zipm}) the linear combination of the
Bessel and Neumann functions. The first one of the coefficients in that
linear combination is obtained from the normalization condition, whereas the
second one is not uniquely determined. In the case $ma<1/2$, the normalized
mode functions are obtained in a way similar to that we have presented
above. They are still given by Eqs. (\ref{psi+}) and (\ref{psi-}) with the
same coefficients $C_{\beta }^{(\pm )}$, where now in Eq. (\ref{Zipm}) we
need to make the replacement%
\begin{equation}
J_{ma\pm 1/2}(x)\rightarrow \frac{J_{ma\pm 1/2}(x)+B_{\beta }Y_{ma\pm 1/2}(x)%
}{\sqrt{1+B_{\beta }^{2}}}.  \label{Irreg}
\end{equation}%
Here, the coefficient $B_{\beta }$ (which, in general, depends on quantum
numbers) should be specified by an additional boundary condition on the AdS
boundary (for a discussion of boundary conditions on fermionic fields in AdS
see Refs. \cite{Brei82,Amse09,Ahar11}). In what follows, for $ma<1/2$ we
consider the boundary condition corresponding to $B_{\beta }=0$. It can be
seen that this corresponds to the situation where the bag boundary condition
is imposed on the fermionic field at $z=\delta >0$ and then the limit $%
\delta \rightarrow 0$ is taken (for the corresponding procedure in the case
of the AdS bulk without compactification see Ref. \cite{Eliz13}). This
ensures the vanishing of the fermionic currents through the AdS boundary.

\section{Fermionic current}

\label{sec:Current}

By using the mode-sum formula (\ref{current}) with the Dirac matrices (\ref%
{gamma}) and the modes (\ref{psi+}), (\ref{psi-}), we can see that the VEVs
of the charge density and of the components of the current density along
noncompact dimensions vanish: $\langle j^{\mu }\rangle =0$, $\mu =0,1,\ldots
,p,D$. For the current density along the $l$th compact dimension, $%
l=p+1,\ldots ,D-1$, we find
\begin{equation}
\langle j^{l}\rangle =-\frac{(4\pi )^{-p/2}eNz^{D+2}}{2\Gamma
(p/2)V_{q}a^{D+1}}\sum_{\mathbf{n}_{q}}\int_{0}^{\infty
}dk_{(p)}\,k_{(p)}^{p-1}\int_{0}^{\infty }d\lambda \,\frac{\lambda k_{l}}{%
\omega }\left[ J_{ma+1/2}^{2}(\lambda z)+J_{ma-1/2}^{2}(\lambda z)\right] ,
\label{jlc}
\end{equation}%
where $\mathbf{n}_{q}=(n_{p+1},\ldots ,n_{D-1})$, $-\infty <n_{i}<+\infty $,
$k_{(p)}^{2}=\sum_{i=1}^{p}k_{i}^{2}$, $\omega ^{2}=\lambda
^{2}+k_{(p)}^{2}+k_{(q)}^{2}$, and%
\begin{equation}
k_{(q)}^{2}=\sum_{i=p+1}^{D-1}k_{i}^{2}=\sum_{i=p+1}^{D-1}\frac{(2\pi n_{i}+%
\tilde{\alpha}_{i})^{2}}{L_{i}^{2}}.  \label{kq2}
\end{equation}%
The same expression for the VEV\ of the current density is obtained in
Appendix \ref{sec:Repr2} by using another representation for the gamma
matrices. In the AdS bulk without compactification the VEV of the current
density vanishes (for a recent discussion of the finite temperature
expectation value see Ref. \cite{Ambr17}).

For the further transformation of the current density, we write the VEV in
the form%
\begin{equation}
\langle j^{l}\rangle =-\frac{(4\pi )^{-p/2}eNz^{D+2}}{2\Gamma
(p/2)V_{q}a^{D+1}}\left[ \mathcal{I}_{ma}(z)+\mathcal{I}_{ma-1}(z)\right] ,
\label{jlc1}
\end{equation}%
with the function%
\begin{equation}
\mathcal{I}_{\nu }(z)=\sum_{\mathbf{n}_{q}}\int_{0}^{\infty
}dk_{(p)}\,k_{(p)}^{p-1}\int_{0}^{\infty }d\lambda \,\frac{\lambda k_{l}}{%
\omega }J_{\nu +1/2}^{2}(\lambda z).  \label{Ical}
\end{equation}%
The transformation of this function is presented in Appendix \ref{sec:App}
with the final result given by Eq. (\ref{Ical2}). As a result, for the
current density we find
\begin{equation}
\langle j^{l}\rangle =-\frac{eNa^{-D-1}L_{l}}{(2\pi )^{(D+1)/2}}%
\sum_{n_{l}=1}^{\infty }n_{l}\sin (\tilde{\alpha}_{l}n_{l})\sum_{\mathbf{n}%
_{q-1}}\,\cos (\tilde{\mathbf{\alpha }}_{q-1}\cdot \mathbf{n}%
_{q-1})\sum_{j=0,1}q_{ma-j}^{\left( D+1\right) /2}(b_{\mathbf{n}_{q}}),
\label{jlc2}
\end{equation}%
with $\mathbf{n}_{q-1}=(n_{p+1},\ldots ,n_{l-1},n_{l+1},\ldots ,n_{D-1})$, $%
\tilde{\mathbf{\alpha }}_{q-1}\cdot \mathbf{n}_{q-1}=\sum_{i=1,\neq l}^{D-1}%
\tilde{\alpha}_{i}n_{i}$, and%
\begin{equation}
b_{\mathbf{n}_{q}}=1+\frac{g_{\mathbf{n}_{q}}^{2}}{2z^{2}},\;g_{\mathbf{n}%
_{q}}^{2}=\sum_{i=p+1}^{D-1}n_{i}^{2}L_{i}^{2}.  \label{b}
\end{equation}%
Here, the function $q_{ma-j}^{\left( D+1\right) /2}(u)$ is defined by Eq. (%
\ref{qf2}) or, equivalently, by the integral representation (\ref{qf}). The
VEV\ of the current density for a charged scalar field is also expressed in
terms of this function \cite{Beze15}.

The VEV of the component of the current density along the compact dimension $%
x^{l}$ is an odd periodic function of $\tilde{\alpha}_{l}$ and an even
periodic function of the remaining phases. In terms of the magnetic fluxes $%
\Phi _{i}$, the period is equal to the flux quantum $\Phi _{0}$. The current
(\ref{jlc2}) determines the charge flux through the spatial hypersurface $%
x^{l}=\mathrm{const}$. Denoting by $n_{i}^{(l)}$, $n_{i}^{(l)}=\delta
_{i}^{l}a/z$, the normal to this hypersurface, for the charge flux one gets
(no summation over $l$) $n_{i}^{(l)}\langle j^{i}\rangle =n_{l}^{(l)}\langle
j^{l}\rangle $. It depends on the lengths $L_{i}$ and on the coordinate $z$
through the ratio $L_{i}/z$. This ratio is the proper length of the $i$th
compact dimension measured by an observer with a fixed $z$, in units of the
curvature radius of the background geometry, $L_{i}/z=L_{(p)i}/a$. We could
expect this feature from the maximal symmetry of the AdS spacetime. Of
course, the current density (\ref{jlc2}) obeys the covariant conservation
equation $\partial _{l}(\sqrt{|g|}\langle j^{l}\rangle )=0$ with $\sqrt{|g|}%
=(a/z)^{D+1}$.

\begin{figure}[tbph]
\begin{center}
\epsfig{figure=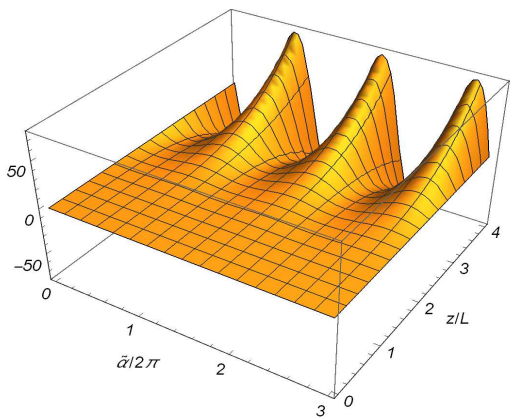,width=7.cm,height=5.5cm}
\end{center}
\caption{The charge flux along the compact dimension, $a^{D}n_{l}^{(l)}%
\langle j^{l}\rangle $, in the model with $p=3$, $q=1$ as a function of the
phase $\tilde{\protect\alpha}$ and of the ratio $z/L$. For the mass we have
taken $ma=1$.}
\label{fig1}
\end{figure}

In the model with a single compact dimension $x^{l}$ with the length $L_{l}=L
$ and with the phase $\tilde{\alpha}_{l}=\tilde{\alpha}$, the general result
(\ref{jlc2}) is reduced to
\begin{equation}
\langle j^{l}\rangle =-\frac{eNa^{-D-1}L}{(2\pi )^{(D+1)/2}}%
\sum_{n=1}^{\infty }n\sin (\tilde{\alpha}n)\sum_{j=0,1}q_{ma-j}^{\left(
D+1\right) /2}\left( 1+\frac{n^{2}L^{2}}{2z^{2}}\right) .  \label{jl1c}
\end{equation}%
In Fig. \ref{fig1}, for $D=4$ and $ma=1$, we have plotted the corresponding
charge flux measured in units of $a^{-D}$, namely, $a^{D}n_{l}^{(l)}\langle
j^{l}\rangle $, as a function of the phase $\tilde{\alpha}$ and of the ratio
$z/L$. The current density is a periodic function of $\tilde{\alpha}$ with
the period $2\pi $. As it will be shown below by asymptotic analysis, the
charge flux vanishes on the AdS boundary and diverges on the AdS horizon.

Let us consider some special cases of the general formula (\ref{jlc2}). For
a massless field, $m=0$, the expressions for the functions $q_{0}^{\left(
D+1\right) /2}(u)$ and $q_{-1}^{\left( D+1\right) /2}(u)$ are most easily
found by using the representation (\ref{qf}) and the expressions for the
functions $I_{\pm 1/2}(x)$. This gives%
\begin{equation}
q_{j}^{\left( D+1\right) /2}(u)=\frac{1}{2}\Gamma \left( \frac{D+1}{2}%
\right) \left[ \frac{1}{\left( u-1\right) ^{(D+1)/2}}-\frac{(-1)^{j}}{\left(
u+1\right) ^{(D+1)/2}}\right] ,  \label{q01}
\end{equation}%
for $j=0,-1$. Plugging into Eq. (\ref{jlc2}), for the current density one
gets $\langle j^{l}\rangle =(z/a)^{D+1}\langle j^{l}\rangle _{\mathrm{M}}$,
where
\begin{equation}
\langle j^{l}\rangle _{\mathrm{M}}=-\frac{e\Gamma ((D+1)/2)}{\pi ^{(D+1)/2}}%
NL_{l}\sum_{n_{l}=1}^{\infty }n_{l}\sin (\tilde{\alpha}_{l}n_{l})\sum_{%
\mathbf{n}_{q-1}}\,\frac{\cos (\tilde{\mathbf{\alpha }}_{q-1}\cdot \mathbf{n}%
_{q-1})}{g_{\mathbf{n}_{q}}^{D+1}},  \label{jlM}
\end{equation}%
is the VEV for a massless fermionic field in $(D+1)$-dimensional Minkowski
spacetime with spatial topology $R^{p+1}\times T^{q}$. The expression (\ref%
{jlM}) is obtained from the more general result from Ref. \cite{Bell10} for
a massive fermionic field in the limit $m\rightarrow 0$ (with the
replacement $\tilde{\alpha}_{l}\rightarrow -2\pi \tilde{\alpha}_{l}$). Note
that, because of the boundary condition we have imposed on the fermionic
modes on the AdS boundary, for a massless field the problem under
consideration is conformally related to the problem in locally Minkowski
spacetime with a boundary at $z=0$ on which the fermionic field obeys the
bag boundary condition. The reason why the current density $\langle
j^{l}\rangle $ is conformally related to the current density $\langle
j^{l}\rangle _{\mathrm{M}}$ in the boundary-free Minkowski case, is that the
boundary-induced contribution in the latter problem vanishes for a massless
field (see Ref. \cite{Bell13}).

\begin{figure}[tbph]
\begin{center}
\epsfig{figure=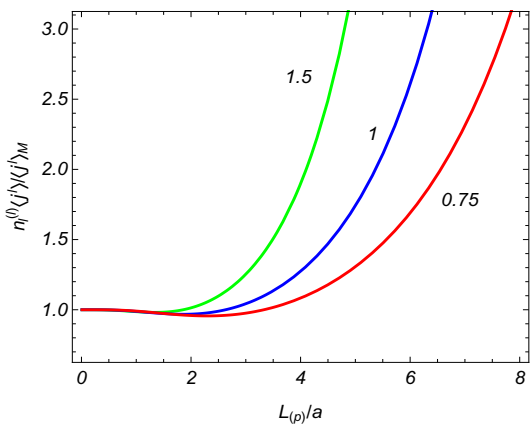,width=7.cm,height=5.5cm}
\end{center}
\caption{The ratio of the charge fluxes in locally AdS and Minkowski
spacetimes, with the same proper lengths of the compact dimension, versus
the proper length. The graphs are plotted for $\tilde{\protect\alpha}=%
\protect\pi /2$ and for different values of $ma$ (numbers near the curves).}
\label{fig2}
\end{figure}

The current density on the Minkowski bulk is obtained in the limit $%
a\rightarrow \infty $ for fixed $y$. The conformal coordinate $z$ is
expanded as $z=a+y+\cdots $. In the integral representation (\ref{qf}) for
the function $q_{\nu }^{\left( D+1\right) /2}(u)$ the order of the Bessel
function is large and we use the corresponding uniform asymptotic expansion.
To the leading order, this gives
\begin{equation}
\sum_{j=0,1}q_{ma-j}^{\left( D+1\right) /2}(u)=2m^{(D+1)/2}a^{D+1}\frac{%
K_{(D+1)/2}(mu)}{u^{(D+1)/2}},  \label{gMink}
\end{equation}%
with $u=b_{\mathbf{n}_{q}}$ and $K_{\nu }(x)$ being the Macdonald function.
Substituting into Eq. (\ref{jlc2}) we get $\lim_{a\rightarrow \infty
}\langle j^{l}\rangle =\langle j^{l}\rangle _{\mathrm{M}}$ with the
Minkowskian result
\begin{equation}
\langle j^{l}\rangle _{\mathrm{M}}=-\frac{2eNL_{l}m^{(D+1)/2}}{(2\pi
)^{(D+1)/2}}\sum_{n_{l}=1}^{\infty }n_{l}\sin (\tilde{\alpha}_{l}n_{l})\sum_{%
\mathbf{n}_{q-1}}\,\cos (\tilde{\mathbf{\alpha }}_{q-1}\cdot \mathbf{n}%
_{q-1})\frac{K_{(D+1)/2}(mg_{\mathbf{n}_{q}})}{g_{\mathbf{n}_{q}}^{(D+1)/2}}.
\label{jlM2}
\end{equation}%
The latter coincides with the expression obtained in Ref. \cite{Bell10}
(again, with the replacement $\tilde{\alpha}_{l}\rightarrow -2\pi \tilde{%
\alpha}_{l}$). In Fig. \ref{fig2}, for the model with $D=4$ and with a
single compact dimension, we have plotted the ratio of the charge fluxes in
locally AdS and Minkowski spacetimes, with the same proper lengths of the
compact dimension, as a function of the proper length measured in units of
the AdS curvature radius $a$. For the phase in the quasiperiodicity
condition along the compact dimension we have taken $\tilde{\alpha}=\pi /2$
and the numbers near the curves correspond to the values of $ma$. As is seen
from the graphs, for large values of the proper length the charge flux in
the AdS bulk is essentially larger than that for the Minkowski case. As it
will be shown below, in the former case the decay of the current density for
large values of the proper length goes like a power law, whereas in the
Minkowski background and for a massive field the decay is exponential (see
Eq. (\ref{jlM2})).

Now we turn to the asymptotics of the current density for small and large
values of $z$. Near the AdS boundary, $z\rightarrow 0$, the argument of the
function $q_{\nu }^{(D+1)/2}(u)$ in Eq. (\ref{jlc2}) is large. By using the
corresponding asymptotic from Ref. \cite{Beze15}, we can see that the
dominant contribution comes from the term $q_{ma-1}^{\left( D+1\right)
/2}(b_{\mathbf{n}_{q}})$ and, to the leading order,
\begin{eqnarray}
\langle j^{l}\rangle &\approx &-\frac{eNL_{l}\Gamma (ma+(D+1)/2)}{\pi
^{D/2}a^{D+1}\Gamma (ma+1/2)}z^{D+1+2ma}  \notag \\
&&\times \sum_{n_{l}=1}^{\infty }n_{l}\sin (\tilde{\alpha}_{l}n_{l})\sum_{%
\mathbf{n}_{q-1}}\,\frac{\cos (\tilde{\mathbf{\alpha }}_{q-1}\cdot \mathbf{n}%
_{q-1})}{g_{\mathbf{n}_{q}}^{D+1+2ma}}.  \label{ljz0}
\end{eqnarray}%
For a massless field this coincides with the exact result. As seen, the
current density vanishes on the AdS boundary as $z^{D+1+2ma}$. For a special
case $D=4$ with a single compact dimension this has been already
demonstrated numerically in figure \ref{fig1}. The large values of $z$
correspond to the near horizon limit. In this limit one has $b_{\mathbf{n}%
_{q}}-1\ll 1$ and by using the asymptotic%
\begin{equation}
q_{\nu }^{\left( D+1\right) /2}(u)\approx \frac{\Gamma \left( (D+1)/2\right)
}{2\left( u-1\right) ^{(D+1)/2}}\left[ 1-\frac{\nu \left( \nu +1\right) }{D-1%
}\left( u-1\right) \right] ,  \label{qnas}
\end{equation}%
valid for $u-1\ll 1$, to the leading order we find $\langle j^{l}\rangle
\approx (z/a)^{D+1}\langle j^{l}\rangle _{\mathrm{M}}$, with $\langle
j^{l}\rangle _{\mathrm{M}}$ given by Eq. (\ref{jlM}). Near the horizon the
dominant contribution comes from the fluctuations with small wavelengths,
and the effects induced by the curvature and nonzero mass are small.

Let us consider the behavior of the current density in asymptotic regions of
the lengths for compact dimensions. If the length of the $l$th compact
dimension is much smaller than the other lengths, $L_{l}\ll L_{i}$, $i\neq l$%
, and $L_{l}\ll z$, the leading contribution comes from the term with $%
\mathbf{n}_{q-1}=0$ and by using the asymptotic expression for the function $%
q_{\nu }^{(D+1)/2}(u)$ for the values of the argument close to 1, one finds%
\begin{equation}
\langle j^{l}\rangle \approx -\frac{eNL_{l}\Gamma ((D+1)/2)}{\pi
^{(D+1)/2}\left( aL_{l}/z\right) ^{D+1}}\sum_{n_{l}=1}^{\infty }\frac{\sin (%
\tilde{\alpha}_{l}n_{l})}{n_{l}^{D}}.  \label{jlLsm}
\end{equation}%
The contribution to the current density from the terms $\mathbf{n}_{q-1}\neq
0$ is suppressed by the exponential factor $e^{-\sigma _{l}|\mathbf{L}\cdot
\mathbf{n}_{q-1}|/L_{l}}$, with $\sigma _{l}=\mathrm{min}(\tilde{\alpha}%
_{l},2\pi -\tilde{\alpha}_{l})$, $0<\tilde{\alpha}_{l}<2\pi $. Note that Eq.
(\ref{jlLsm}) coincides with the current density for a massless field in the
model with a single compact dimension $x^{l}$.

For large values of the proper length of the $l$th compact dimension
compared with the AdS curvature radius one has $L_{(p)l}/a=L_{l}/z\gg 1$
and, hence, $b_{\mathbf{n}_{q}}\gg 1$. By using the asymptotic expression
for the function $q_{\nu }^{(D+1)/2}(u)$ for large values of the argument,
for the leading order term in the current density one gets the result (\ref%
{ljz0}). If in addition $L_{l}\gg L_{i}$, $i\neq l$, the contribution from
large values of $|n_{i}|$ dominates and the corresponding summations can be
replaced by the integration. Two cases should be considered separately. If
all the phases $\tilde{\alpha}_{i}$, $i\neq l$, are zero the leading term is
given by
\begin{equation}
\langle j^{l}\rangle \approx -\frac{Ne\Gamma (p/2+ma+3/2)}{\pi
^{p/2+1}\Gamma (ma+1/2)a^{D+1}V_{q}}\frac{z^{D+2ma+1}}{L_{l}^{p+2ma+1}}%
\sum_{n_{l}=1}^{\infty }\frac{\sin (\tilde{\alpha}_{l}n_{l})}{n_{l}^{p+2ma+2}%
}.  \label{jllargeL}
\end{equation}%
As is seen, in this case, for a given $z$, the decay of the current density
as a function of the proper length $L_{(p)l}$ follow a power law. For a
massive field this behavior essentially differs from that for the current
density in the Minkowski bulk. In the latter geometry and for a massive
field, the current decays exponentially, as $e^{-mL_{l}}$. Provided at least
one of the phases $\tilde{\alpha}_{i}$, $i\neq l$, is different from zero,
the current density is dominated by the term $n_{l}=1$ and one finds%
\begin{equation}
\langle j^{l}\rangle \approx -\frac{Nea^{-1-D}\sin (\tilde{\alpha}%
_{l})z^{D+2ma+1}}{2\pi ^{(p+1)/2}\Gamma (ma+1/2)V_{q}}\frac{\kappa
^{p/2+ma+1}e^{-\kappa L_{l}}}{(2L_{l})^{p/2+ma}},  \label{jllargeL2}
\end{equation}%
with the notation $\kappa ^{2}=\sum_{i=p+1,\neq l}^{D-1}\tilde{\alpha}%
_{i}^{2}/L_{i}^{2}$. Now we have an exponential decay as a function of $%
L_{l} $.

\begin{figure}[tbph]
\begin{center}
\epsfig{figure=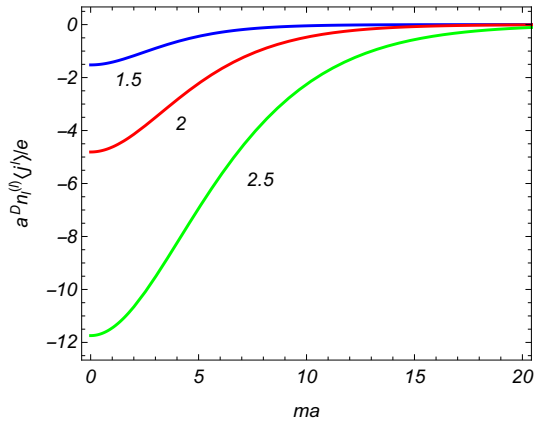,width=7.cm,height=5.5cm}
\end{center}
\caption{The charge flux along the compact dimension, $a^{D}n_{l}^{(l)}%
\langle j^{l}\rangle $, as a function of the field mass for different values
of the ratio $z/L$ (numbers near the curves). The graphs are plotted for $%
\tilde{\protect\alpha}=\protect\pi /2$.}
\label{fig3}
\end{figure}

For large values of the mass, $ma\gg 1$, the dominant contribution to the
series in the right-hand side of Eq. (\ref{jlc2}) comes from the term with $%
n_{i}=0$, $i\neq l$, and $n_{l}=1$. Introducing the notation $%
u=L_{l}/(2z)=L_{(p)l}/(2a)$, for the leading order contribution one finds%
\begin{equation}
\langle j^{l}\rangle \approx -\frac{eNL_{l}\sin (\tilde{\alpha}_{l})}{2(4\pi
)^{D/2}a^{D+1}}\frac{(ma)^{D/2}u^{-D/2-1}(1+u^{2})^{-D/4}}{(1+2u^{2}+2u\sqrt{%
1+u^{2}})^{ma+1/2}}(u+\sqrt{1+u^{2}}),  \label{jlargem}
\end{equation}%
and the VEV is exponentially suppressed. In Fig. \ref{fig3}, the dependence
of the charge flux along the compact dimension on the mass of the fermionic
field is displayed for the background with $D=4$ with a single compact
dimension. The graphs are plotted for $\tilde{\alpha}=\pi /2$ and the
numbers near the curves are the corresponding values of the ratio $z/L$.

From the covariant conservation equation for the current density it follows
that the charge flux through the hypersurface element $d^{D-1}x=dz\prod%
\limits_{i=1,\neq l}^{D-1}dx^{i}$ is given by $\sqrt{|g|}j^{l}d^{D-1}x$. For
the expectation value of the total charge flux, per unit coordinate surface
along spatial dimensions $(x^{1},\ldots ,x^{l-1},x^{l+1},\ldots ,x^{D-1})$,
integrated over $z$, one gets $\int_{0}^{\infty }dz\,\left( a/z\right)
^{D+1}\langle j^{l}\rangle $. From the asymptotic analysis of the current
density near the AdS boundary and horizon, given above, we can see that the
integral is convergent in the lower limit and linearly diverges in the upper
limit. Hence, similar to the Minkowskian case, the total charge flux
diverges.

Comparing the fermionic current density in the Minkowski bulk, Eq. (\ref%
{jlM2}), with the corresponding expression from Ref. \cite{Beze13c} for the
current density $\langle j^{l}\rangle _{\mathrm{M}}^{\mathrm{(s)}}$ of a
charged scalar field, we see that%
\begin{equation}
\langle j^{l}\rangle _{\mathrm{M}}=-(N/2)\langle j^{l}\rangle _{\mathrm{M}}^{%
\mathrm{(s)}},  \label{RelMink}
\end{equation}%
provided the masses, charges and the phases in the periodicity conditions
are the same for fermionic and scalar fields. In particular, in
supersymmetric models with the same number of fermionic and scalar degrees
of freedom, the total current in the Minkowski bulk vanishes. This is not
the case for the AdS bulk (for a discussion of boundary conditions on the
AdS boundary in supersymmetric models see, for example, Refs. \cite%
{Amse09,Ahar11} and references therein). Note that for supersymmetric models
in AdS background the fields in the same multiplet do not necessarily have
the same mass (see, e.g., Refs. \cite{Amse09,Wit99}). By taking into account
the expression of the current density for scalar fields \cite{Beze15}, for
the total current in the system of a fermionic field and $N/2$ charged
scalar fields (equal number of fermionic and scalar degrees of freedom) one
gets%
\begin{eqnarray}
\langle j^{l}\rangle ^{\mathrm{(t)}} &=&\frac{Nea^{-1-D}L_{l}}{(2\pi
)^{(D+1)/2}}\sum_{n_{l}=1}^{\infty }n_{l}\sin (\tilde{\alpha}_{l}n_{l})\sum_{%
\mathbf{n}_{q-1}}\,\cos (\tilde{\mathbf{\alpha }}_{q-1}\cdot \mathbf{n}%
_{q-1})  \notag \\
&&\times \left[ 2q_{\nu _{s}-1/2}^{(D+1)/2}(b_{\mathbf{n}_{q}})-q_{ma}^{%
\left( D+1\right) /2}(b_{\mathbf{n}_{q}})-q_{ma-1}^{\left( D+1\right) /2}(b_{%
\mathbf{n}_{q}})\right] ,  \label{jltot}
\end{eqnarray}%
where $\nu _{s}=\sqrt{D^{2}/4-D(D+1)\xi +m^{2}a^{2}}$ and $\xi $ is the
curvature coupling parameter for scalar fields. Near the AdS horizon, the
leading contribution from the scalar and fermionic parts in Eq. (\ref{jltot}%
) cancel each other and we need to keep the next-to-leading term in the
asymptotic (\ref{qnas}). This leads to the result%
\begin{eqnarray}
\langle j^{l}\rangle ^{\mathrm{(t)}} &\approx &\frac{D(D+1)NeL_{l}}{4\pi
^{(D+1)/2}a^{D+1}}\Gamma \left( \frac{D-1}{2}\right) \left( \xi -\xi
_{D}\right) z^{D-1}  \notag \\
&&\times \sum_{n_{l}=1}^{\infty }n_{l}\sin (\tilde{\alpha}_{l}n_{l})\sum_{%
\mathbf{n}_{q-1}}\,\frac{\cos (\tilde{\mathbf{\alpha }}_{q-1}\cdot \mathbf{n}%
_{q-1})}{g_{\mathbf{n}_{q}}^{D-1}},  \label{jltotas}
\end{eqnarray}%
in the limit $z\rightarrow \infty $. Here, $\xi _{D}=(D-1)/(4D)$ is the
value of the curvature coupling parameter for conformal coupling. The
leading term, given by Eq. (\ref{jltotas}), does not depend on the mass. For
$ma<1/2$ or $ma\geqslant 1/2$ and $D(D+1)\left( \xi -\xi _{D}\right) <ma$,
the fermionic part dominates near the AdS boundary and the total VEV $%
\langle j^{l}\rangle ^{\mathrm{(t)}}$ behaves as in Eq. (\ref{ljz0}). In
particular, the latter is the case for minimally coupled scalar fields. For $%
ma\geqslant 1/2$ and $D(D+1)\left( \xi -\xi _{D}\right) >ma$, the total
current density\ near the AdS boundary is dominated by the scalar
contribution and $\langle j^{l}\rangle ^{\mathrm{(t)}}\propto z^{D+2\nu
_{s}+2}$ for $z\rightarrow 0$.

\begin{figure}[tbph]
\begin{center}
\begin{tabular}{cc}
\epsfig{figure=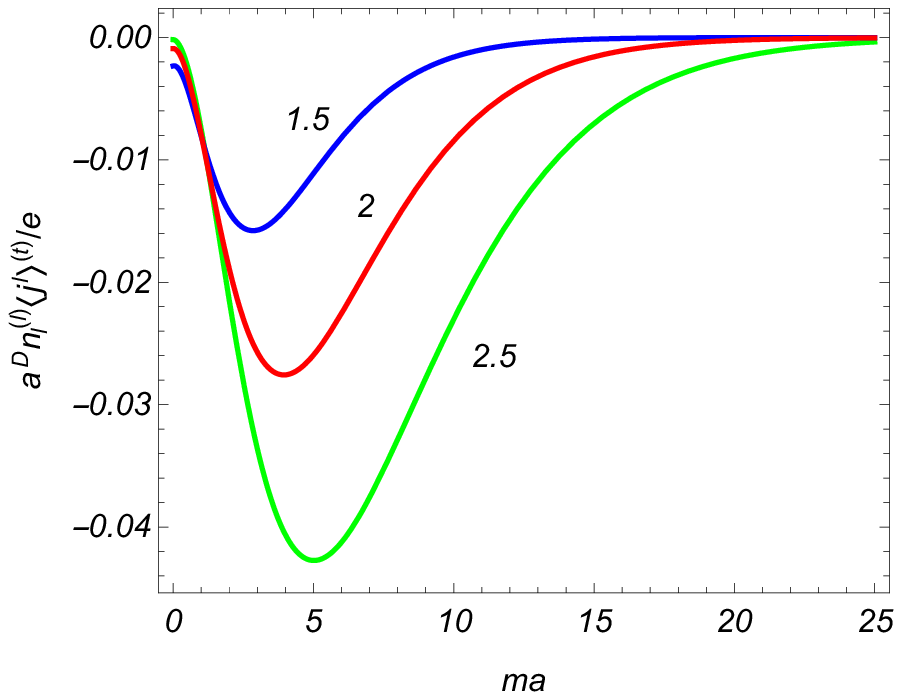,width=7.cm,height=5.5cm} & \quad %
\epsfig{figure=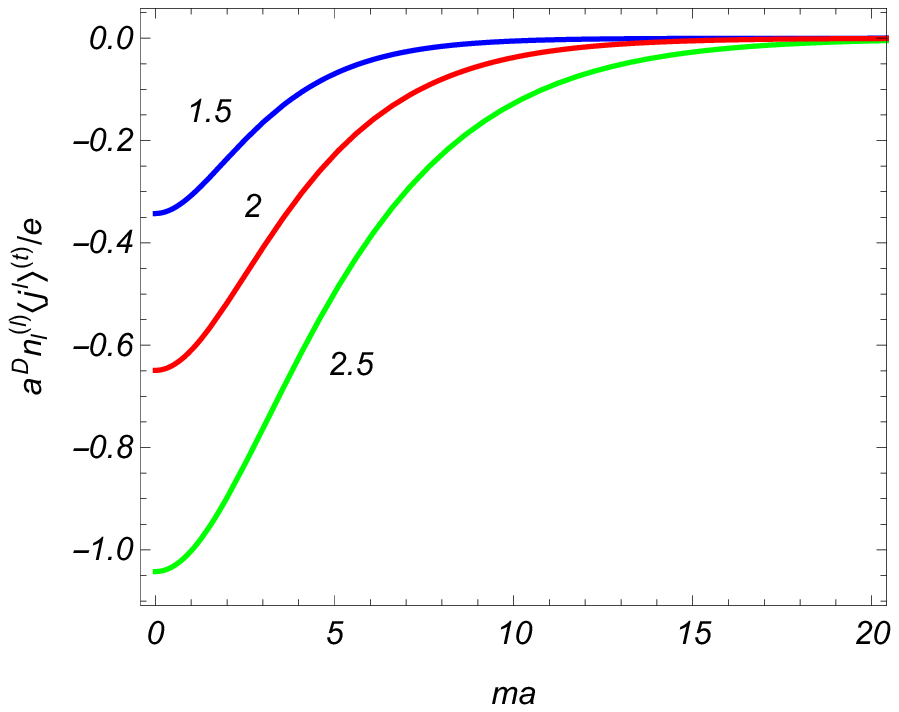,width=7.cm,height=5.5cm}%
\end{tabular}%
\end{center}
\caption{The total charge flux in $(p,q)=(3,1)$ model with a fermionic field
and $N/2$ scalar fields as a function of the mass. The left/right panels
correspond to conformally/minimally coupled scalar fields. The graphs are
plotted for $\tilde{\protect\alpha}=\protect\pi /2$ and the numbers near the
curves correspond to the values of the ratio $z/L$.}
\label{fig4}
\end{figure}

In Fig. \ref{fig4}, the dependence on the mass of the total charge flux is
plotted in the $(p,q)=(3,1)$ model with a fermionic field and $N/2$ scalar
fields. The graphs are plotted for $\tilde{\alpha}=\pi /2$ and the numbers
near the curves correspond to the values of the ratio $z/L$. The left and
right panels are for conformally and minimally coupled scalar fields. In
both cases the total current is dominated by the fermionic contribution. In
general, the current density is not a monotonic function of the mass.

As seen from Eq. (\ref{jltotas}), compared to the separate scalar and
fermionic contributions, the divergence of the total current density on the
horizon is weaker and, as a consequence of that, the charge flux integrated
over the $z$-coordinate is finite. By using in Eq. (\ref{jltot}) the
integral representation (\ref{qf}) for the function $q_{\nu }^{\left(
D+1\right) /2}(u)$, after the evaluation of the integral over $z$, the $x$%
-integral is reduced to $\int_{0}^{\infty }dx\,e^{-x}g(x)$ with $%
g(x)=\sum_{j=\pm 1/2}I_{ma+j}(x)-2I_{\nu _{s}}(x)$. Though this integral is
convergent, the separate integrals with the modified Bessel functions
diverge. In order to have the right to take the integrals separately, we can
replace the integral by $\int_{0}^{\infty }dx\,e^{-bx}g(x)$, $b>1$. After
the evaluation of the separate integrals by using the formula from Ref. \cite%
{Prud86}, the limit $b\rightarrow 1$ is taken easily. In this way, for the
integrated charge flux we get%
\begin{equation}
\int_{0}^{\infty }dz\,\left( \frac{a}{z}\right) ^{D+1}\langle j^{l}\rangle ^{%
\mathrm{(t)}}=\left( ma-\nu _{s}\right) \frac{\Gamma (D/2)eNL_{l}}{2(4\pi
)^{D/2}}\sum_{n_{l}=1}^{\infty }n_{l}\sin (\tilde{\alpha}_{l}n_{l})\sum_{%
\mathbf{n}_{q-1}}\,\frac{\cos (\tilde{\mathbf{\alpha }}_{q-1}\cdot \mathbf{n}%
_{q-1})}{g_{\mathbf{n}_{q}}^{D}}.  \label{chargeflt}
\end{equation}%
It is of interest to note that the mass enters in the factor $\left( ma-\nu
_{s}\right) $ only.

\section{Fermionic current in C-,P- and T-symmetric odd-dimensional models
on AdS bulk}

\label{sec:OddDim}

In the discussion above we have evaluated the VEV of the fermionic current
density for a field realizing the irreducible representation of the Clifford
algebra. It is known that (see, for example, Ref. \cite{Shim85}) in even
values of the spatial dimension $D$ (odd-dimensional spacetimes) the mass
term $m\bar{\psi}\psi $ in the Lagrangian density breaks $P$-invariance, $C$%
-invariance for $D=4n$, and $T$-invariance in $D=4n+2$ with $n=0,1,2,\ldots $%
. $P$-, $C$-, and $T$-invariant models can be constructed combining two
fermionic fields in the irreducible representations. In this section we
assume that the flat spacetime Dirac matrices are taken in the
representation (\ref{gam2}). For odd-dimensional spacetimes the matrix $%
\gamma ^{(D)}$ can be expressed in terms of the product $\gamma =\gamma
^{(0)}\gamma ^{(1)}\cdots \gamma ^{(D-1)}$ in two inequivalent ways, namely,
$\gamma ^{(D)}=\gamma _{(s)}^{(D)}=s\gamma $, $s=\pm 1$, for $D=4n$ and $%
\gamma ^{(D)}=\gamma _{(s)}^{(D)}=si\gamma $ for $D=4n+2$. The upper and
lower signs correspond to two inequivalent representations of the
corresponding Clifford algebra with the gamma matrices $\gamma _{(s)}^{(\mu
)}=(\gamma ^{(0)},\gamma ^{(1)},\cdots \gamma ^{(D-1)},\gamma _{(s)}^{(D)})$%
. The corresponding matrices in AdS spacetime can be taken as $\gamma
_{(s)}^{\mu }=(a/z)\gamma _{(s)}^{(\mu )}$.

Consider a system of two $N$-component fermionic fields, $\psi _{(+1)}$ and $%
\psi _{(-1)}$, with the combined Lagrangian density $\mathcal{L}=\sum_{s=\pm
1}\bar{\psi}_{(s)}(i\gamma _{(s)}^{\mu }\nabla _{\mu }^{(s)}-m)\psi _{(s)}$,
$\nabla _{\mu }^{(s)}=\partial _{\mu }+\Gamma _{\mu }^{(s)}$. By suitable
transformations of the fields (in general, mixing the separate fields) it
can be seen that the Lagrangian density is invariant under the $C$-, $P$-
and $T$-transformations (see, e.g., Ref. \cite{Shim85}). By taking into
account the relations $\gamma ^{(0)}\gamma ^{\dagger }\gamma ^{(0)}\gamma
=-1 $ and $\gamma ^{(0)}\gamma ^{\dagger }\gamma ^{(0)}\gamma _{(+1)}^{\mu
}\gamma =\gamma _{(-1)}^{\mu }$, the Lagrangian density can also be
rewritten as
\begin{equation}
\mathcal{L}=\sum_{s=\pm 1}\bar{\psi}_{(s)}^{\prime }(i\gamma ^{\mu }\nabla
_{\mu }-sm)\psi _{(s)}^{\prime }.  \label{Lodd}
\end{equation}%
with $\psi _{(+1)}^{\prime }=\psi _{(+1)}$, $\psi _{(-1)}^{\prime }=\gamma
\psi _{(-1)}$, and $\gamma ^{\mu }=\gamma _{(+1)}^{\mu }$. As seen, the new
field $\psi _{(-1)}^{\prime }$ satisfies the same equation as $\psi _{(+1)}$
with the opposite sign for the mass term. Introducing $2N\times 2N$ Dirac
matrices $\gamma ^{(2N)\mu }=\sigma _{\mathrm{P}3}\otimes \gamma ^{\mu }$,
with $\sigma _{\mathrm{P}3}=\mathrm{diag}(1,-1)$ being the Pauli matrix, the
Lagrangian density is presented in terms of the $2N$-component spinor $\Psi
=(\psi _{(+1)}^{\prime },\psi _{(-1)}^{\prime })^{T}$ in the form
\begin{equation}
\mathcal{L}=\bar{\Psi}[i\gamma ^{(2N)\mu }(\partial _{\mu }+\Gamma _{\mu
}^{(2N)})-m]\Psi ,  \label{Loddb}
\end{equation}%
where $\Gamma _{\mu }^{(2N)}=I\otimes \Gamma _{\mu }$.

In the system of two fermionic fields realizing two inequivalent
representations of the Clifford algebra, the total current density $J^{\mu
}=e\bar{\Psi}\gamma ^{(2N)\mu }\Psi $ is given by $J^{\mu }=\sum_{s=\pm
1}j_{(s)}^{\mu }$, where $j_{(s)}^{\mu }=e\bar{\psi}_{(s)}\gamma _{(s)}^{\mu
}\psi _{(s)}$ are the current densities for separate fields. As it has been
shown in Appendix \ref{sec:Repr2}, if the phases in the quasiperiodicity
conditions along compact dimensions are the same for the fields $\psi
_{(+1)} $ and $\psi _{(-1)}$ then the VEVs of the corresponding current
densities are the same. In this case, the VEV $\langle J^{l}\rangle $ is
obtained from the expressions for $\langle j^{l}\rangle $ given above with
an additional coefficient 2. However, the phases $\alpha _{l}$ for the
fields $\psi _{(+1)} $ and $\psi _{(-1)}$, in general, can be different.

Among the most interesting physical realizations of fermionic systems in
odd-dimensional spacetimes is the graphene. Graphene is an one-atom thick
layer of graphite. The low-energy excitations of the corresponding electron
subsystem can be described by a pair of two-component spinors, composed of
the Bloch states residing on the two triangular sublattices $A$ and $B$ of
the graphene~honeycomb lattice. For the spatial dimension in the
corresponding effective field theory one has $D=2$. For a given value of
spin $S=\pm 1$, two spinors are combined in a 4-component spinor $\Psi
_{S}=(\psi _{+,AS},\psi _{+,BS},\psi _{-,AS},\psi _{-,BS})^{T}$. The
components $\psi _{\pm ,AS}$ and $\psi _{\pm ,BS}$ correspond to the
amplitude of the electron wave function on sublattices $A$ and $B$ and the
indices $+$ and $-$ correspond to inequivalent points, $\mathbf{K}_{+}$ and $%
\mathbf{K}_{-}$ of the Brillouin zone (see Refs. \cite{Ando05,Gusy07}). The
values of the parameter $s=+1$ and $s=-1$ in the discussion above,
specifying the irreducible representations of the Clifford algebra,
correspond to these points. Consequently, we can identify $\psi _{(s)}=(\psi
_{s,AS},\psi _{s,BS})^{T}$. For the bulk geometry with vanishing 0th
component of the spin connection, $\Gamma _{0}=0$, and for the zero scalar
potential of the external electromagnetic field (these conditions are the
case in the problem at hand) the combined Lagrangian density, in standard
units, is presented as%
\begin{equation}
L=\sum_{S=\pm 1}\bar{\Psi}_{S}(i\hbar \gamma ^{(4)0}\partial _{t}+i\hbar
v_{F}\gamma ^{(4)l}D_{l}-\Delta )\Psi _{S}.  \label{LagGr}
\end{equation}%
Here, $v_{F}\approx 7.9\times 10^{7}$ cm/s is the Fermi velocity of the
electrons in graphene, $D_{l}=(\mathbf{\nabla }-ie\mathbf{A}/\hbar c)_{l}$,
with $l=1,2$, is the spatial part of the gauge extended covariant
derivative, and for electrons $e=-|e|$. The energy gap $\Delta $ can be
created by a number of mechanisms (see, for instance, Ref. \cite{Gusy07} and
references therein) and it is related to the Dirac mass $m$ through $\Delta
=mv_{F}^{2}$. For the analog of the Compton wavelength corresponding to the
energy gap one has $a_{\mathrm{C}}=\hbar v_{F}/\Delta $. The energy scale in
the system is determined by the combination $\gamma _{F}=\hbar v_{F}/a_{0}$ (%
$\approx 2.51$ eV), where $a_{0}\approx 1.42$ \AA\ is the inter-atomic
spacing of the graphene lattice. In terms of this combination one has $a_{%
\mathrm{C}}=a_{0}\gamma _{F}/\Delta $. The Lagrangian densities (\ref{LagGr}%
) for separate $S$ are the analog of Eq. (\ref{Loddb}) for $D=2$.

For a planar graphene sheet the spatial topology in the corresponding
effective field theory is $R^{2}$. In the case of a sheet rolled into a
cylinder (cylindrical carbon nanotubes) or torus (toroidal nanotubes) the
topology becomes nontrivial: $R^{1}\times S^{1}$ and $S^{1}\times S^{1}$,
respectively. In these systems, the magnetic fluxes $\Phi _{l}$ ($l=2$ for
cylindrical nanotubes and $l=1,2$ for toroidal ones) we have introduced
above, acquire real physical meaning. The carbon nanotubes are characterized
by chiral vector $\mathbf{C}_{h}=n_{a}\mathbf{a}+n_{b}\mathbf{b}$, where $%
n_{a}$, $n_{b}$ are integers and $\mathbf{a}=(\sqrt{3},0)a_{0}$, $\mathbf{b}%
=(-\sqrt{3},3)a_{0}/2$ are primitive translation vectors of the graphene
hexagonal lattice (for general properties of carbon nanotubes, see, for
example, Ref. \cite{Sait98}). Note that $\sqrt{3}a_{0}$ is the lattice
constant. In the construction of the nanotube the hexagon at the origin is
identified with the hexagon at $\mathbf{C}_{h}$. For zigzag and armchair
nanotubes one has $n_{b}=0$ and $n_{a}=2n_{b}$, respectively. The other
pairs $(n_{a},n_{b})$ correspond to chiral nanotubes. In the case $%
n_{a}+n_{b}=3q_{c}$, $q_{c}\in Z$, the nanotube will be metallic and for $%
n_{a}+n_{b}\neq 3q_{c}$ the nanotube will be semiconducting. In the latter
case, the corresponding energy gap is inversely proportional to the
diameter. In particular, the armchair nanotube is metallic and the $%
(n_{a},0) $ zigzag nanotube is metallic if and only if $n_{a}$ is an integer
multiple of 3. The chirality also determines the periodicity condition along
the compact dimension for the fields $\psi _{(s)}$. In the absence of the
magnetic flux threading the nanotube, the periodicity conditions have the
form $\psi _{(s)}(\mathbf{r}+\mathbf{C}_{h})=e^{-2s\pi ip_{c}/3}\psi _{(s)}(%
\mathbf{r}+\mathbf{C}_{h})$, where, for a given nanotube, the parameter $%
p_{c}=-1,0,+1$ is defined by the relation $n_{a}+n_{b}=3q_{c}+p_{c}$ (see,
e.g., the discussion in Ref. \cite{Ando05}). Hence, for the phase we have
introduced before, one has $\alpha =-2s\pi ip_{c}/3$. For metallic nanotubes
one has the periodic boundary condition, $\alpha =0$, and for semiconducting
ones $\alpha =\pm 2\pi /3$. As seen, the phases for the spinors
corresponding to the points, $\mathbf{K}_{+}$ and $\mathbf{K}_{-}$ of the
Brillouin zone have opposite signs. As a consequence, in the absence of the
magnetic flux, the total current density in cylindrical nanotubes vanishes.

\begin{figure}[tbph]
\begin{center}
\epsfig{figure=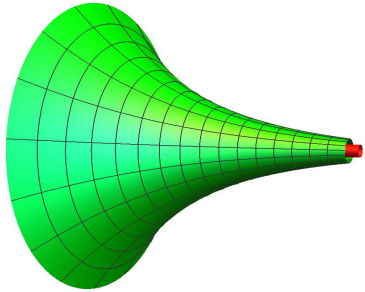,width=6.cm,height=5.5cm}
\end{center}
\caption{The spatial geometry of a $D=2$ tube embedded in 3-dimensional
Euclidean space. The tube of the magnetic flux is shown as well.}
\label{fig5}
\end{figure}

The problem under consideration is topologically equivalent to the case of
cylindrical nanotubes, though the corresponding spatial geometry is curved
(for the generation of curvature in graphene sheets and the corresponding
effects on the properties of graphene see, for example, Ref. \cite%
{Kole09,Iori13}). In Fig. \ref{fig5} we have plotted this geometry embedded
into a 3-dimensional Euclidean space. The magnetic flux threading the
compact dimension is shown as well. The proper length of the compact
dimension is decreasing with increasing $z$. Note that, related to the
graphene physics, a similar spatial geometry has been discussed in \cite%
{Iori13}. However, the spacetime geometry we consider here is different.

\begin{figure}[tbph]
\begin{center}
\begin{tabular}{cc}
\epsfig{figure=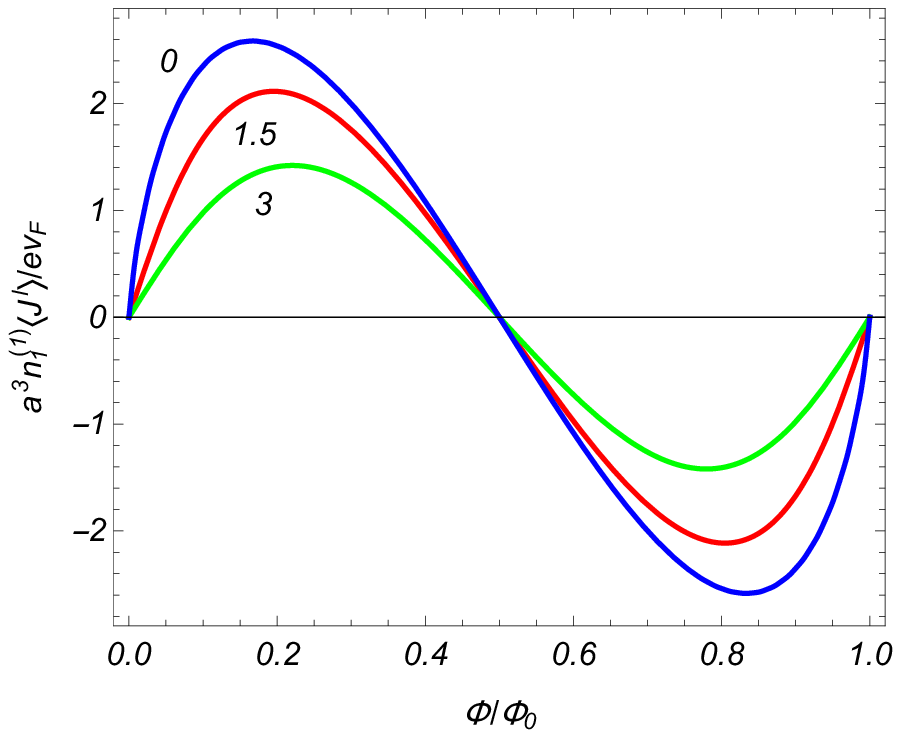,width=7.cm,height=5.5cm} & \quad %
\epsfig{figure=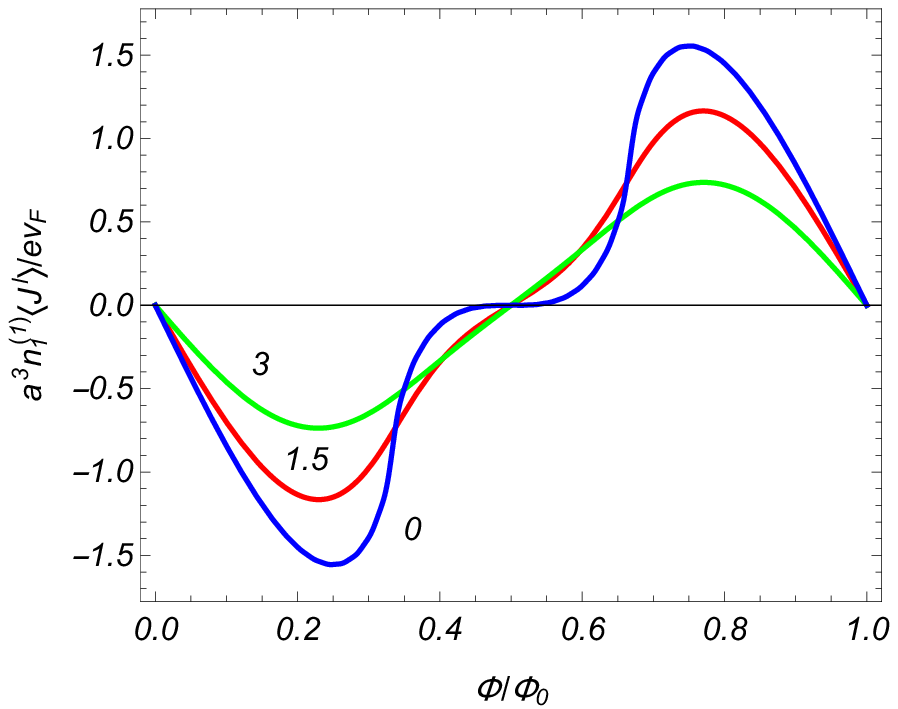,width=7.cm,height=5.5cm}%
\end{tabular}%
\end{center}
\caption{The total charge flux in $D=2$ tubes as a function of the magnetic
flux for separate values of $a/a_{\mathrm{C}}$ (numbers near the curves).
The left/right panels correspond to metallic/semiconducting tubes. The
graphs are plotted for $z/L=2$.}
\label{fig6}
\end{figure}

Consequently, for a given $S$, the VEVs of the current densities for
separate contributions coming from the points $\mathbf{K}_{+}$ and $\mathbf{K%
}_{-}$ are given by the expressions in previous sections with an additional
factor $v_{F}$, and for the nonzero component of the total current one has $%
\langle J^{1}\rangle =\sum_{s=\pm 1}\langle j_{(s)}^{1}\rangle $, where $%
j_{(\pm 1)}^{1}$ are the contributions from two valleys. In the problem
under consideration separate spins $S$ give the same contributions in the
ground state currents. Assuming that the phases $\alpha $ for the
contributions from $s=+1$ and $s=-1$ have opposite signs, one finds%
\begin{equation}
\langle J^{1}\rangle =-\frac{\sqrt{2}ev_{F}L}{\pi a^{3}}\sum_{n=1}^{\infty
}n\cos (\alpha _{(+1)}n)\sin \left( 2\pi n\frac{\Phi }{\Phi _{0}}\right)
h\left( a/a_{\mathrm{C}},1+\frac{n^{2}L^{2}}{2z^{2}}\right) ,  \label{J1cn}
\end{equation}%
with the function%
\begin{equation}
h\left( \nu ,x\right) =2^{\nu }\partial _{x}\left[ \frac{1/\sqrt{x-1}}{%
\left( \sqrt{x+1}+\sqrt{x-1}\right) ^{2\nu }}\right] .  \label{hxy}
\end{equation}%
In the absence of the magnetic flux the current density vanishes. In Eq. (%
\ref{J1cn}), the ratio $a/a_{\mathrm{C}}$ is the analog of the product $ma$
in the discussion of the previous sections. If the curvature of the tube
does not change the phases, then $\alpha _{(+1)}=0$ for metallic tubes and $%
\alpha _{(+1)}=2\pi /3$ for semiconducting ones. Fig. \ref{fig6} presents
the current density in these two cases versus the magnetic flux treading the
tube for different values of $a/a_{\mathrm{C}}$ (numbers near the curves).
The left/right panels correspond to the metallic/semiconducting tubes. In
the numerical evaluation we have taken $z/L=2$.

\section{Conclusion}

\label{sec:Conc}

In the present paper we have investigated the VEV\ of the fermionic current
density in $(D+1)$-dimensional AdS spacetime with a toroidally compactified
subspace. The periodicity conditions for the field operator along compact
dimensions contain arbitrary phases $\alpha _{l}$ and, in addition, the
presence of a constant abelian gauge field is assumed. By a gauge
transformation the problem is reduced to the one in the absence of the gauge
field with the shifted phases (\ref{alftild}) in the quasiperiodicity
conditions for the new field. The phase shift for the $l$th compact
dimension is formally interpreted in terms of the magnetic flux enclosed by
that dimension.

For the evaluation of the vacuum currents we have used the direct summation
over the complete set of fermionic modes (\ref{psi+}) and (\ref{psi-}). The
same result is obtained by using the alternative set of fermionic mode
functions (\ref{psi+n}) and (\ref{psi-n}). The VEVs of the charge density
and of the components for the current density along uncompact dimensions
vanish and the mode sum for the component of the current density along the $%
l $th compact dimension is presented as Eq. (\ref{jlc}). A more convenient
representation for the renormalized VEV is given by Eq. (\ref{jlc2}). The
current density along the $l$th compact dimension is an even periodic
function of the phases $\tilde{\alpha}_{i}$, $i\neq l$, and an odd periodic
function of the phase $\tilde{\alpha}_{l}$. In particular, this means the
periodicity in the magnetic flux with the period equal to the flux quantum.
The current density is expressed in terms of the function (\ref{qf2}). An
alternative integral representation is given by Eq. (\ref{qf}). Note that
the VEV\ of the current density for a charged scalar field is also expressed
through the function (\ref{qf2}).

In order to clarify the behavior of the current density, various limiting
cases of the general result are considered. First of all, we have shown
that, in the limit of the infinite curvature radius, the current density for
a locally Minkowski bulk is obtained. For a massless field the problem under
consideration is conformal to the one in Minkowski spacetime with toral
dimensions in the presence of a boundary on which the fermionic field
operator obeys the bag boundary condition. However, the boundary-induced
contribution in the latter problem vanishes for a massless field and we have
obtained a conformal relation with the boundary-free Minkowski case. On the
AdS boundary the VEV\ of the current density vanishes as $z^{D+1+2ma}$,
whereas on the AdS horizon it diverges as $z^{D+1}$. Near the horizon the
dominant contribution comes from the fluctuations with small wavelengths and
the effects induced by the curvature and nonzero mass are small. The
influence of the curvature on the component of the current density along the
$l$th compact dimension is also small, when the corresponding length $L_{l}$
is much smaller than other length scales in the problem. The leading term in
the corresponding asymptotic expansion coincides with the VEV for a massless
field in the model with a single compact dimension $x^{l}$. In the opposite
limit of large values for the proper length of the $l$th compact dimension,
the behavior of the VEV $\langle j^{l}\rangle $ is essentially different
depending on the values of the phases in the quasiperiodicity conditions. If
all the remaining phases vanish, $\tilde{\alpha}_{i}=0$, $i\neq l$, the
current density, as a function of the proper length, behaves as $%
1/L_{(p)l}^{p+2ma+1}$. Unlike the case of the locally Minkowski bulk, the
corresponding decay is power law for both massless and massive fields. If at
least one of the phases $\tilde{\alpha}_{i}$, $i\neq l$, is different from
zero, the decay of the VEV\ $\langle j^{l}\rangle $, as a function of $L_{l}$%
, is exponential.

In the Minkowski bulk, for the system of a fermionic field and $N/2$ charged
scalar fields, with the same masses, charges and the same phases in the
quasiperiodicity conditions, the total current vanishes as a consequence of
the cancellation between the fermionic and scalar contributions. This means
that, in the corresponding supersymmetric models, no net current appears. In
the AdS bulk the influence of the gravitational field on the VEVs for scalar
and fermionic fields, in general, is different and there is no cancellation
between the scalar and fermionic counterparts. The corresponding total
current density is given by Eq. (\ref{jltot}). Near the horizon, the leading
contributions from the scalar and fermionic parts are canceled, and the
first term in the corresponding asymptotic expansion is presented as Eq. (%
\ref{jltotas}). It does not depend on the mass and vanishes for conformally
coupled scalars. As a consequence of the weaker divergence on the horizon,
the total charge flux integrated over the $z$-coordinate is finite.

In odd spacetime dimensions the mass term in the Lagrangian density for a
fermionic field realizing the irreducible representation of the Clifford
algebra, in general, breaks $C$-, $P$-, and $T$-invariances. The models with
these symmetries can be constructed combining two fermionic fields realizing
two irreducible representations. In section \ref{sec:OddDim} we have
considered the current density in this type of models. It has been shown
that, if the phases $\alpha _{l}$ are the same for both the representations,
their contributions to the total current coincide. However, the phases need
not be the same. This type of situation arises in semiconducting cylindrical
carbon nanotubes described by an effective Dirac theory in the
long-wavelength approximation. In the corresponding Dirac model two
irreducible representations correspond to two inequivalent points of the
graphene Brillouin zone and, in the absence of the magnetic flux threading
the tube, the corresponding phases have opposite signs. We have considered
the fermionic current in the corresponding problem on the AdS bulk generated
by a magnetic flux.

\section*{Acknowledgments}

A. A. S. was supported by the State Committee of Science Ministry of
Education and Science RA, within the frame of Grant No. SCS 15T-1C110. The
work was partially supported by the NATO Science for Peace Program under
Grant No. SFP 984537. V. V. acknowledges support through De Sitter cosmology
fellowship. A. A. S. gratefully acknowledges the hospitality of the INFN,
Laboratori Nazionali di Frascati (Frascati, Italy), where a part of this
work was done.

\appendix

\section{Another class of fermionic modes}

\label{sec:Repr2}

In the evaluation of the current density we have used the representation (%
\ref{gamma}) for the Dirac matrices and the corresponding fermionic modes (%
\ref{psi+}), (\ref{psi-}). As it has been discussed in Ref. \cite{Eliz13},
these modes are well adapted for the investigation of the effects induced by
the presence of an additional brane, parallel to the AdS boundary, on which
the field operator obeys the bag boundary condition. In this appendix we
consider another representation of the Dirac matrices that allows for the
separation of the equations for the upper and lower components of the
fermionic mode functions. In the new representation the flat spacetime gamma
matrices are given by
\begin{equation}
\gamma ^{(0)}=\left(
\begin{array}{cc}
0 & \chi _{0} \\
\chi _{0}^{\dagger } & 0%
\end{array}%
\right) ,\;\gamma ^{(l)}=\left(
\begin{array}{cc}
0 & \chi _{l} \\
-\chi _{l}^{\dagger } & 0%
\end{array}%
\right) ,  \label{gam2}
\end{equation}%
where $l=1,2,\ldots ,D-1$, and $\gamma ^{(D)}=si\,\mathrm{diag}(1,-1)$ with $%
s=\pm 1$. In odd-dimensional spacetimes, the values $s=+1$ and $s=-1$
correspond to two irreducible representations of the Clifford algebra. For
the $N/2\times N/2$ matrices $\chi _{0}$, $\chi _{l}$ one gets the relations
$\chi _{l}\chi _{n}^{\dagger }+\chi _{n}\chi _{l}^{\dagger }=2\delta _{nl}$,
$\chi _{l}^{\dagger }\chi _{n}+\chi _{n}^{\dagger }\chi _{l}=2\delta _{nl}$
for $l,n=1,2,\ldots ,D-1$, and $\chi _{0}\chi _{l}^{\dagger }=\chi _{l}\chi
_{0}^{\dagger }$, $\chi _{0}^{\dagger }\chi _{l}=\chi _{l}^{\dagger }\chi
_{0}$, $\chi _{0}^{\dagger }\chi _{0}=1$. As before, the curved spacetime
gamma matrices are given by $\gamma ^{\mu }=(z/a)\delta _{b}^{\mu }\gamma
^{(b)}$. In the special case $D=2$ we have $\chi _{0}=\chi _{1}=1$ and $%
\gamma ^{(0)}=\sigma _{\mathrm{P}1}$, $\gamma ^{(1)}=i\sigma _{\mathrm{P}2}$%
, $\gamma ^{(2)}=i\sigma _{\mathrm{P}3}$, where $\sigma _{\mathrm{P}\mu }$
are the Pauli matrices (see, for instance, Ref. \cite{Park09}).

For the positive-energy modes, decomposing the spinor into the upper and
lower components, the separate equations are obtained for them with the
solution
\begin{equation}
\psi _{\beta }^{(+)}=z^{\frac{D+1}{2}}e^{i\mathbf{kx}-i\omega t}\left(
\begin{array}{c}
J_{ma+s/2}(\lambda z)\chi ^{(\sigma )} \\
\frac{\mathbf{k\chi }^{\dagger }+\omega \chi _{0}^{\dagger }}{\lambda }%
J_{ma-s/2}(\lambda z)\chi ^{(\sigma )}%
\end{array}%
\right) ,  \label{psi+2}
\end{equation}%
where $\chi ^{(\sigma )}$, $\sigma =$ $1,\ldots ,N/2$, are one-column
matrices with $N/2$ rows and $\mathbf{k\chi }^{\dagger
}=\sum_{l=1}^{D-1}k_{l}\chi _{l}^{\dagger }$. From the orthonormalization
condition for the modes (\ref{psi+2}) one obtains%
\begin{equation}
\chi ^{(\sigma )\dagger }[\left( \omega +\mathbf{k\chi }\chi _{0}^{\dagger
}\right) ^{2}+\lambda ^{2}]\chi ^{(\sigma ^{\prime })}=2\lambda
^{2}\left\vert C_{\beta }^{(+)}\right\vert ^{2}\delta _{\sigma \sigma
^{\prime }},  \label{ortxi}
\end{equation}%
where $\left\vert C_{\beta }^{(+)}\right\vert ^{2}$ is given by Eq. (\ref%
{Cbet}). This shows that we can take
\begin{equation}
\left( \omega +\mathbf{k\chi }\chi _{0}^{\dagger }-i\lambda \right) \chi
^{(\sigma )}=\sqrt{2}\lambda C_{\beta }^{(+)}w^{(\sigma )},  \label{mat1}
\end{equation}%
or inverting%
\begin{equation}
\chi ^{(\sigma )}=C_{\beta }^{(+)}\frac{\mathbf{k\chi }\chi _{0}^{\dagger
}+i\lambda -\omega }{\sqrt{2}i\omega }w^{(\sigma )},  \label{mat2}
\end{equation}%
where the matrices $w^{(\sigma )}$ are the same as in Eq. (\ref{psi+}).

As a result, for the positive-energy fermionic modes we get
\begin{equation}
\psi _{\beta }^{(+)}=\frac{C_{\beta }^{(+)}}{\sqrt{2}}z^{\frac{D+1}{2}}e^{i%
\mathbf{kx}-i\omega t}\left(
\begin{array}{c}
\frac{\mathbf{k\chi }\chi _{0}^{\dagger }+i\lambda -\omega }{\omega }%
J_{ma+s/2}(\lambda z)w^{(\sigma )} \\
i\chi _{0}^{\dagger }\frac{\mathbf{k\chi }\chi _{0}^{\dagger }+i\lambda
+\omega }{\omega }J_{ma-s/2}(\lambda z)w^{(\sigma )}%
\end{array}%
\right) .  \label{psi+n}
\end{equation}%
In a similar way, for the negative-energy modes one finds the representation%
\begin{equation}
\psi _{\beta }^{(-)}=\frac{C_{\beta }^{(-)}}{\sqrt{2}}z^{\frac{D+1}{2}}e^{i%
\mathbf{kx+}i\omega t}\left(
\begin{array}{c}
i\chi _{0}\frac{\mathbf{k\chi }^{\dagger }\chi _{0}-i\lambda +\omega }{%
\omega }J_{ma+s/2}(\lambda z)w^{(\sigma )} \\
\frac{\mathbf{k\chi }^{\dagger }\chi _{0}-i\lambda -\omega }{\omega }%
J_{ma-s/2}(\lambda z)w^{(\sigma )}%
\end{array}%
\right) .  \label{psi-n}
\end{equation}%
with $C_{\beta }^{(-)}$ defined in Eq. (\ref{Cbet}). As an additional check,
it can be seen that the modes (\ref{psi+n}) and (\ref{psi-n}) are orthogonal.

With the modes (\ref{psi+n}) and (\ref{psi-n}), we can evaluate the
fermionic current density by using the mode-sum formula (\ref{current}). By
taking into account that for a $N/2\times N/2$ matrix $M$ we have $%
\sum_{\sigma }w^{(\sigma )\dagger }Mw^{(\sigma )}=\mathrm{tr\,}M$, the VEV\
of the component of the current density along the compact direction $x^{l}$
is presented as%
\begin{equation}
\langle j^{l}\rangle =-\frac{eNz^{D+2}}{4(2\pi )^{p}V_{q}a^{D+1}}\sum_{%
\mathbf{n}_{q}}\int d\mathbf{k}_{(p)}\,\int_{0}^{\infty }d\lambda \frac{%
k_{l}\lambda }{\omega }\sum_{j=\pm 1}J_{ma+j/2}^{2}(\lambda z).
\label{jl2nd}
\end{equation}%
After the integration over the angular part of $\mathbf{k}_{(p)}$ this
expression is reduced to Eq. (\ref{jlc}). As seen, the VEV\ of the current
density does not depend on the parameter $s$ in the definition of the Dirac
matrix $\gamma ^{(D)}$. In particular, in odd-dimensional spacetimes the
current density is the same for two inequivalent irreducible representations
of the Clifford algebra. From the derivation of the expression (\ref{jl2nd})
it follows that it is also valid for $D=2$ model with the compact dimension
of the length $L$. The corresponding expression is obtained putting in Eq. (%
\ref{jl2nd}) $N=2$, $p=0$, $V_{q}=L$, and omitting the integral over $%
\mathbf{k}_{(p)}$.

\section{Evaluation of the mode-sum}

\label{sec:App}

Here we evaluate the mode-sum in the definition (\ref{Ical}) for the
function $\mathcal{I}_{\nu }(z)$. By using the integral representation%
\begin{equation}
\frac{1}{\omega }=\frac{1}{\sqrt{\pi }}\int_{0}^{\infty }\frac{ds}{s^{1/2}}%
\,e^{-\omega ^{2}s},  \label{rep}
\end{equation}%
the integral over $\lambda $ in Eq. (\ref{Ical}) is expressed in terms of
the modified Bessel function $I_{\nu }(z^{2}/2s)$ \cite{Prud86}. After the
integration over $k_{(p)}$ one finds%
\begin{equation}
\mathcal{I}_{\nu }(z)=\frac{\Gamma (p/2)}{4\sqrt{\pi }}\int_{0}^{\infty }%
\frac{ds}{s^{(p+3)/2}}e^{-z^{2}/(2s)}I_{\nu +1/2}(z^{2}/2s)\sum_{\mathbf{n}%
_{q}}k_{l}e^{-k_{(q)}^{2}s}.  \label{Ical1}
\end{equation}%
As the next step we employ the relation%
\begin{equation}
\sum_{n_{j}=-\infty }^{+\infty }e^{-sk_{j}^{2}}=\frac{L_{j}}{2\sqrt{\pi s}}%
\sum_{n_{j}=-\infty }^{+\infty }e^{in_{j}\tilde{\alpha}%
_{j}}e^{-L_{j}^{2}n_{j}^{2}/(4s)},  \label{Pois2}
\end{equation}%
which is a direct consequence of the Poisson resummation formula. From Eq. (%
\ref{Pois2}) we can see that%
\begin{equation}
\sum_{\mathbf{n}_{q}}k_{l}e^{-k_{(q)}^{2}s}=\frac{-iL_{l}V_{q}}{2^{q+1}\pi
^{q/2}s^{q/2+1}}\sum_{\mathbf{n}_{q}}n_{l}e^{i\tilde{\mathbf{\alpha }}\cdot
\mathbf{n}_{q}-g_{\mathbf{n}_{q}}^{2}/(4s)},  \label{Pois3}
\end{equation}%
with $\tilde{\mathbf{\alpha }}\cdot \mathbf{n}_{q}=$ $\sum_{i=p+1}^{D-1}n_{i}%
\tilde{\alpha _{i}}$ and $g_{\mathbf{n}_{q}}$ is defined by Eq. (\ref{b}).

With these relations, the function (\ref{Ical}) is presented in the form%
\begin{equation}
\mathcal{I}_{\nu }(z)=\frac{-iL_{l}V_{q}\Gamma (p/2)}{2^{q-(D-3)/2}\pi
^{q/2+1}z^{D+2}}\sum_{\mathbf{n}_{q}}n_{l}e^{i\tilde{\mathbf{\alpha }}\cdot
\mathbf{n}_{q}}q_{\nu }^{(D+1)/2}(b_{\mathbf{n}_{q}}),  \label{Ical2}
\end{equation}%
where $b_{\mathbf{n}_{q}}$ is given by Eq. (\ref{b}). Here we have defined
the function%
\begin{equation}
q_{\nu }^{\left( D+1\right) /2}(u)=\sqrt{\frac{\pi }{2}}\int_{0}^{\infty
}dx\,x^{D/2}e^{-ux}I_{\nu +1/2}(x).  \label{qf}
\end{equation}%
The integral is expressed in terms of the hypergeometric function and one
gets an alternative representation%
\begin{equation}
q_{\nu }^{(D+1)/2}(u)=\frac{\sqrt{\pi }\Gamma (\nu +D/2+3/2)}{2^{\nu
+1}\Gamma (\nu +3/2)u^{\nu +D/2+3/2}}F\left( \frac{2\nu +D+3}{4},\frac{2\nu
+D+5}{4};\nu +\frac{3}{2};\frac{1}{u^{2}}\right) .  \label{qf2}
\end{equation}%
From the definition (\ref{qf}) it directly follows that $q_{\nu }^{\left(
D+1\right) /2}(u)$ is a monotonically decreasing function of $u>1$ and $\nu
>-1/2$. For even number of spatial dimensions, $D=2n$, $n=1,2,\ldots $, the
expression (\ref{qf2}) for the function $q_{\nu }^{\left( D+1\right) /2}(u)$
is further simplified to \cite{Beze15}
\begin{equation}
q_{\nu }^{n+1/2}(u)=(-1)^{n}\sqrt{\frac{\pi }{2}}\partial _{u}^{n}\frac{(u+%
\sqrt{u^{2}-1})^{-\nu -1/2}}{\sqrt{u^{2}-1}}.  \label{qev}
\end{equation}%
In odd dimensional space, $D=2n-1$, the function (\ref{qf2}) is expressed in
terms of the Legendre function of the second kind: $q_{\nu
}^{n}(u)=(-1)^{n}\partial _{u}^{n}Q_{\nu }(u)$.

\end{document}